\documentstyle[12pt,aasms4]{article}
\received{August 18, 1999}
\accepted{October 28, 1999}
\lefthead{Tomisaka}
\righthead{Evolution of Angular Momentum Distribution during
 Star Formation}
\begin{document}

\title{The Evolution of the Angular Momentum Distribution during
 Star Formation}
\author{Kohji Tomisaka\altaffilmark{1}}
\affil{Faculty of Education and Human Sciences, 
 Niigata University, 8050 Ikarashi-2, Niigata 950-2181}
\altaffiltext{1}{E-mail: tomisaka@ed.niigata-u.ac.jp}

\begin{abstract}
If the angular momentum of the molecular cloud core were conserved
 during the star formation process, a new-born star would rotate much
 faster than its fission speed.  
This constitutes the angular momentum problem of new-born stars.  
In this paper, the angular momentum transfer in the contraction of a
 rotating magnetized cloud is studied with axisymmetric MHD simulations.  
Owing to the large dynamic range covered by the nested-grid method,
 the structure of the cloud in the range from 10 AU to 0.1 pc is explored.  
First, the cloud experiences a run-away collapse, and a disk forms
 perpendicularly to the magnetic field, in which the central density
 increases greatly in a finite time-scale.  
In this phase, the specific angular momentum $j$ of the disk decreases
 to $\simeq 1/3$ of the initial cloud.  
After the central density of the disk exceeds $\sim 10^{10}{\rm cm}^{-3}$,
 the infall on to the central object develops.  
In this accretion stage, the rotation motion and thus the toroidal
 magnetic field drive the outflow.  
The angular momentum of the central object is transferred efficiently
 by the outflow as well as the effect of the magnetic stress.  
In 7000 yr from the core formation, the specific angular momentum of 
 the central $0.17M_\odot$ decreases a factor of $10^{-4}$ from the 
 initial value (i.e. from $10^{20}{\rm cm^2~s^{-1}}$ to 
 $10^{16}{\rm cm^2~s^{-1}}$). 
\end{abstract}


\section{INTRODUCTION}

The problem of star formation has been the object of serious study for
many years.  In the star formation process, almost all the angular
momentum should be removed from the molecular cloud in the course of
forming stars.  For example, using rotation periods and the radius of
classical T Tauri stars (CTTS) [$P=3-10$ days and $R_* \simeq 2
R_\odot$ (Bouvier et al. 1993)], the specific angular momentum of such
stars is approximately equal to $j\simeq 4\pi R_*^2/(5P) \sim 5.6
\times 10^{16} {\rm cm^2~s}^{-1}(R_*/2 R_\odot)^2(P/10{\rm
day})^{-1}$, where we assumed the star rotates uniformly.  On the
other hand, if the velocity gradient of $0.3{\rm km~s^{- 1}~pc^{-1}}-4
{\rm km~s^{-1}~pc^{-1}}$ (Goodman et al. 1993) observed in ${\rm
NH_3}$ cores (density $\gtrsim 10^4{\rm cm}^{-3}$; size $\sim 0.1$ pc)
comes from the rotation, $j$ reaches $\sim 5\times 10^{21} {\rm
cm^2~s}^{-1} (R/0.1{\rm pc})^2(\Omega/4 {\rm km~s^{-1}~pc^{- 1}})$.
Therefore, the specific angular momentum should decrease a factor of
$10^{- 5}$ when a star is formed from a molecular cloud core.
Assuming $j$ is conserved through the collapse, the gravity is
balanced with the centrifugal force at the centrifugal radius as
$R_c=j^2/(GM)\sim 0.06 {\rm pc}(j/5\times 10^{21} {\rm cm^2~s}^{-1})^2
(M/1M_\odot)^{-1}$.  This leads to the conclusion that if the angular
momentum is not efficiently reduced or transferred, the cloud could
not shrink to become a stellar object ($R_c \gg R_*$).  This is called
the angular momentum problem for new-born stars.
 
To reduce the spin angular momentum of newly formed stars, two
mechanisms are proposed.  One is to transfer the cloud spin angular
momentum to the orbital angular momentum of binary or multiple
protostars if they are made by fission.  The other possibility is
magnetic braking, in which angular momentum is transferred to the
ambient medium by the magnetic stress (Spitzer 1978).  The latter
works even for the case of a single star.  In the case of the parallel
rotator, in which the directions of magnetic fields and angular
momentum coincide with each other, the time scale of deceleration of
the angular rotation speed ($\Omega$) is given by \begin{equation} t_B
\equiv -\frac{\Omega}{d\Omega/dt} \simeq \frac{\sigma}{2\rho_a
V_A}=\frac{2\pi G \sigma}{B_0}(4\pi G \rho_a)^{-1/2}, \end{equation}
where $\sigma$, $\rho_a$, $V_A$, and $G$ denote, respectively, the
column density of the molecular core, the ambient density, the
Alfv\'{e}n speed of the ambient medium, and the gravitational constant
(Ebert, von Hoerner, \& Temesvary 1960, Mouschovias, \& Paleologou
1980).  As long as the molecular core evolves in a quasistatic manner,
the decrease in the angular momentum $L$ is well fitted by the above
time scale as $L=L(t=0)\exp(-t/t_B)$ (Tomisaka, Ikeuchi, \& Nakamura
1990).  Since the first factor ${2\pi G \sigma}/{B_0}$ is
approximately equal to or larger than unity for a gravitationally
contracting cloud (Tomisaka 1995), $t_B$ is longer than the free-fall
time of the ambient low-density medium.  This raises the question of
whether or not the angular momentum is efficiently transferred while
the contraction proceeds.

Recently, bipolar outflows have been found in a wide variety of
objects which indicate star formation (for a review, see Bachiller
1996).  Based on the magnetic acceleration model of outflow
(Blandford, \& Payne 1982; see also K\"{o}nigl 1989 and Wardle, \&
K\"{o}nigl 1993), magnetic fields extract the angular momentum from
the disk and transfer it to the outflow gas.  Therefore, the outflow
gas is able to carry the excess angular momentum directly.

{}From MHD simulations of the gravitational contraction of a
 magnetized, rotating isothermal molecular cloud,
 it is shown that the outflow begins after the central density
 exceeds $10^{10}{\rm cm}^{-3}$ and an adiabatic core forms in
 the center (Tomisaka 1998, hereafter Paper I).  
The outflow is never driven in the run-away collapse phase,
 since rotation motion and thus toroidal magnetic fields are
 unimportant in this phase.  
Therefore, in this paper, we study the evolution of the cloud from 
 the runaway collapse phase to the accretion phase, continuously.  
In particular, we try to determine when the redistribution of angular
 momentum occurs and whether the outflow plays an important role in
 extracting the angular momentum from the contracting gas.  
The model and the numerical method are described in $\S$2.  
In $\S$ 3, the evolution of the specific angular momentum distribution
 is shown.  
In addition, a comparison is made between the effects of the magnetic
 torque and outflow to reduce the angular momentum.  
We discuss the ideal MHD condition and the evolution of late stage in $\S$4.

\section{Model and Numerical Method}

We began our simulation from an infinitely long, cylindrical,
 rotating, isothermal cloud in hydrostatic balance.  
In terms of the gravitational potential, $\psi_0$, the magnetic
 flux density, {\bf B}$_0=B_z${\bf e}$_z$, the angular rotation speed,
 $\Omega_0$, and the density $\rho_0$,
 the hydrostatic configuration is governed by the equations 
\begin{equation} 
 r\Omega_0^2-\frac{d \psi_0}{d r}
 -\frac{c_s^2}{\rho_0}\frac{d\rho_0}{dr}
 -\frac{1}{8\pi}\frac{dB_z^2}{dr}=0, 
\end{equation} 
\begin{equation}
 \frac{1}{r}\frac{d}{dr}\left(r\frac{d\psi_0}{dr}\right)=4\pi G\rho_0.
\end{equation} 
These equations have a solution for the radial distributions of density,
 rotational velocity, and magnetic flux density as follows
 (Stod\'{o}\l{}kiewicz 1963): 
\begin{equation}
 \rho_0(r)=\rho_{c}\left[1+r^2/(8H^2)\right]^{-2}, 
\end{equation}
\begin{equation}
 v_\phi(r)\equiv r\Omega_0(r)=r\Omega_{c}\left[1+r^2/(8H^2)\right]^{-1/2},
\end{equation} 
\begin{equation}
 B_z(r)=B_{c}\left[1+r^2/(8H^2)\right]^{-1}, 
\end{equation}
where $H$ represents the scale-height as
 $4\pi G \rho_c H^2=c_s^2+B_c^2 / (8\pi\rho_c)+2 \Omega_c^2 H^2$
 [for $\Omega_c < (2\pi G \rho_c)^{1/2}$].  
Here, quantities with a subscript c denote those for the cloud center
 ($r=0$).  
This hydrostatic solution contains two nondimensional parameters:
 $\omega\equiv\Omega_c/(4\pi G \rho_c)^{1/2}$ and 
 $\alpha\equiv B_{c}^2/(4\pi\rho_c c_s^2)$.

We assume that the gas obeys the ideal MHD equations.  
The gas is assumed isothermal for low densities.  
However, after the density exceeds $\sim 10^{10}{\rm cm}^{-3}$,
 the interstellar gas is not efficiently cooled and behaves
 adiabatically (Larson 1969).  
To mimic this situation, a barotropic relation such as 
\begin{equation}
 p \left\{ 
 \begin{array}{l} =c_s^2 \rho, \ldots (\rho < \rho_{\rm crit})\\
                  =c_s^2 \rho_{\rm crit} (\rho/\rho_{\rm crit})^\Gamma
                                  , \ldots (\rho > \rho_{\rm crit})
 \end{array}\right.  
\end{equation} 
 was adopted without using the energy equation,
 where we take $\rho_{\rm crit}= 10^{10} {\rm H_2}{\rm cm}^{-3}$.  
The polytropic exponent $\Gamma$ was taken as 5/3 and 2.  
(The specific heat ratio $\gamma$ never exceeds 5/3.  
However, we found that the flow outside the core
 and the mass-accretion rate are essentially the same irrespective 
 of the values of the core's polytropic exponent, $\Gamma$.  
Therefore, we used the larger exponent to reduce the effort to perform
 numerical computations.)  
To initiate gravitational contraction, we added density perturbation
 with small amplitude.  
The periodic boundary condition was applied to the boundaries in the
 $z$-direction, and the separation between two boundaries was chosen to
 be identical with the wavelength of the gravitationally most unstable
 mode (Matsumoto, Hanawa, \& Nakamura 1997).  
To ensure the fine spatial resolution necessary especially near the forming
 adiabatic core, the nested-grid method was applied to the MHD finite
 difference scheme (Tomisaka 1996a, 1996b, 1998).  
In the nested-grid method, 15 levels of grids are used from L0
 (the coarsest) to L14 (the finest).  
A grid spacing of L$n$ was chosen as 1/2 of that of L$n-1$, and each
 level of grids was divided into $64\times 64$ cells.  
The code was tested by comparing the result of gravitational collapse
 calculated by a code without the nested-grid method (for details,
 see Tomisaka 1996b).

\section{RESULTS}

Here, the results of the model with $\omega=1/2$ and $\alpha=1$ are shown.  
This corresponds to $\Omega=2.78n_{c\,4}^{1/2}\,{\rm km~s^{-1}~pc^{-1}}$
 and $B_c=13.3 n_{c\,4}^{1/2} c_{s\,190}\,\mu{\rm G}$.  
Here, $n_{c\,4}$ and $c_{s\,190}$ represent respectively
 $\rho_c/10^4 {\rm H_2~cm}^{-3}$ and $c_s/190{\rm m~s}^{-1}$.  
The parameters are taken to be the same as the model shown in Paper I.
Evolutions are also the same.

\subsection{Specific Angular Momentum Distribution}

In Figure 1, the specific angular momentum $j$ is plotted against the
 mass measured from the center.  
Since the density increases monotonically reaching the center,
 a mass shell with higher density is considered to be located nearer
 to the center than that with lower density.  
The total mass of the gas whose density is higher than $\rho_1$, 
\begin{equation} 
 M(>\rho_1)=\int_{\rho > \rho_{1}}\rho (z,r) dV,
\end{equation}
and the correspondent total angular momentum,
\begin{equation}
 L(>\rho_1)\equiv\int_{\rho > \rho_{1}}\rho (z,r) v_\phi r dV,
\end{equation}
are calculated at three epochs.  
Figure 1 is made by plotting $j(>\rho_1)\equiv L/M$ against $M(>\rho_1)$
 for various $\rho_1$.  
This is similar to the specific angular momentum spectrum (e.g. Fig.5 of
 Norman, Wilson, \& Barton 1980) in which $M(<j)$, the total mass whose 
 specific angular momentum is smaller than $j$ is plotted against $j$.
However, the plot in Figure 1 has an advantage that the gas with smaller $M$ is
 necessarily located nearer to the center.

Open circles represent the $j(<M)$ distribution in the early phase 
[ $t=5.214(4\pi G \rho_c)^{-1/2} \sim 0.9n_{c\,4}^{-1/2} {\rm Myr}$ ],
 in which the central density increases a factor 10 from the initial state
 (i.e., $\rho_c$ reaches $\simeq 10^5 n_{c\,4}{\rm H_2\,cm^{-3}}$).  
At this stage, the density distribution is almost spherical.  
Filled squares denote the distribution at the end of the run-away collapse
 phase ($t<5.980(4\pi G \rho_c)^{-1/2}\sim 1.05 n_{c\,4}^{-1/2}{\rm Myr} $;
 the structure at the stage is shown in Fig. 2a of Paper I).  
Since the motion crossing the magnetic fields is blocked, a disk running
 perpendicularly to the magnetic fields is formed.  
From Figure 1, it is shown that the angular momentum contained in the central
 $\sim 0.5 c_{s\,190}^3 n_{c\,4}^{-1/2}M_\odot$
 is reduced to $\simeq 1/3$ before a core is formed. 

This reduction is not owing to the angular momentum transfer by the magnetic
 fields but to the change of the shape of high-density region (from spherical
 to disk-like).
This is confirmed by the fact that the specific angular momentum spectrum
 (not shown in this Letter) is not changed during the run-away collapse phase.
The disk is mainly formed by the gas flow parallel to the
 magnetic fields ($z$-direction).  
Therefore, central part of the run-away collapsing disk is made by a gas
 with small angular momentum sitting initially near the $z$-axis.
In other words, to the high-density region 
 the mass is preferentially gathered rather than the angular momentum.
The difference between open circles and filled squares in Figure 1
 is owing to the segregation of low-angular momentum gas in the run-away
 collapse stage.
However, the amount of the angular momentum segregation is too
 insufficient to explain the angular momentum problem.
  
The specific angular momentum, $j$, is approximately proportional to
 the accumulated mass, $M$.  
This relation coincides with the prediction for the thin disk by Basu(1998).  
This relation is explained from the solution of the contracting
 isothermal thin disk, i.e., the column density $\sigma\simeq c_s/Gr$
 and the angular rotation speed $\Omega \propto r^{-1}$ (Matsumoto et al.
 1997) lead to the specific angular momentum 
 $j=\Omega r^2\propto r\propto M\equiv \int \sigma r dr$.

After the central density of the disk exceeds
 $\rho_{\rm crit}\sim 10^{10}{\rm H_2 cm^{-3}}$,
 a new phase appears: the accretion phase.
Thermal photons emitted from the dust which cooled the cloud become
 optically thick; 
as a result, an almost spherical adiabatic core is formed 
 (in the adiabatic core, the thermal pressure becomes more important).  
Gas around the core begins to accrete on the core.

As shown in Paper I, after $\sim 10^3$ yr has passed,\footnote{
 This age is independent on the values of the assumed polytropic index,
  $\Gamma=5/3$ and 2.
 Furthermore, the mass outflow rate differs slightly: $\Gamma=2$ model shows
  approximately 20 \% smaller outflow rate than the model of $\Gamma=5/3$.
 However, in the case of slowly rotating cloud as $\alpha=1$ and
  $\omega=1/10$, which leads to a smaller centrifugal radius, 
  in a model assuming $\Gamma=2$  the outflow begins earlier  than
  than that of $\Gamma=5/3$.
 This seems to indicate that unless the centrifugal radius from which the outflow
  is mainly ejected is formed near the center, especially, in the hard core,
  assuming such a relatively hard polytropic index is justified.
 Quantitative comparison will be made in a forthcoming paper.}  
 the outflow begins (Figures 2b and 3 of Paper I).  
The size of the seeding region (origin of the outflow) is
 $r\sim 5\times 10^{-3}c_s/(4\pi G \rho_c)^{1/2} 
 \sim 35 c_{s\,190}n_{c\,4}^{-1/2}{\rm AU}$.  
The seeding region expands radially outward.  
This outflow is driven by the gradient of the magnetic pressure of
 the toroidal magnetic fields, $-\nabla B_\phi^2/8\pi$, which are 
 made by the rotation motion (Paper I;Kudoh, Matsumoto, \& Shibata 1998).  
The magnetic fields exert torque on the outflowing gas to increase its
 angular momentum.  
On the other hand, they exert torque on the disk to decrease the angular
 momentum (Blandford, \& Payne 1982).  
In $\tau=0.04(4\pi G \rho_c)^{-1/2}\sim 7000 n_{c\,4}^{-1/2}{\rm yr}$ 
 (filled triangles; $\tau$ is the time elapsed from the core formation epoch),
 the outflow expands and reaches $z\sim 0.25 c_s(4\pi G \rho_c)^{-1/2}\sim
 1800 c_{s\,190}n_{c\,4}^{-1/2}{\rm AU}$.
The angular momentum distribution at that time is shown by the solid line
 in Figure 1.
It is shown that the specific angular momentum contained in the central
 $\sim 0.17 c_{s\,190}^3 n_{c\,4}^{-1/2}M_\odot$ ($\rho > 
 10^{10}n_{c\,4}{\rm H_2~cm}^{-3}$) has been
 reduced to a factor $\lesssim 10^{-4}$ of the initial value.

\subsection{Angular Momentum Flux and Magnetic Torques}

Consider the total angular momentum, $L(<M)$, contained in the mass $M$.  
The increase and decrease of the angular momentum are caused by
 (1) the inflow (accretion flow) mainly in the disk $dL/dt|_{\rm in}$,
 (2) the torque exerted on the gas disk from the magnetic fields $N$,
 and (3) the outflow from the disk $dL/dt|_{\rm out}$:
 $dL/dt=dL/dt|_{\rm in}+dL/dt|_{\rm out}+N$.  
It is to be noted that only the first term is positive.  
To see which term is important to reduce the angular momentum,
 we plot these three quantities for the three epochs in Figure 2.  
To calculate $dL/dt|_{\rm in}$ and $dL/dt|_{\rm out}$, we consider a
 cylinder which covers the isodensity surface.  
The $dL/dt|_{\rm out}$ is measured by the angular momentum flux convected
 outwardly through the upper and lower surfaces of the cylinder,
 and $dL/dt|_{\rm in}$ is the flux running radially inward through
 the side surface.

Before the core formation (circles and squares), the angular momentum
 inflow ($dL/dt|_{\rm in}$) is larger than the magnetic torque ($N$)
 which reduces the angular momentum.  
(This does not mean that the specific angular momentum increases.  
Note that $dM/dt_{\rm in}$ is also positive.)  
In contrast, in the accretion phase after outflow blows,
 the magnetic torque (filled triangles)
 is comparable with $dL/dt|_{\rm in}$ (open triangles) near 
 $M \gtrsim 0.17 c_{s\,190}^3 n_{c\,4}^{-1/2}M_\odot$ 
 ($\tau=7000 n_{c\,4}^{-1/2}{\rm yr}$).
Further, from the figure, it is shown that the angular momentum
 transport by outflow $dL/dt|_{\rm out}$ (open stars) is more efficient
 than that of the magnetic torque.  
By these two effects, the angular momentum transferred by the inflow
 inside the disk is totally removed for central 
 $\sim 0.17 c_{s\,190}^3 n_{c\,4}^{-1/2} M_\odot$.  
As a result, $j(<0.17 c_{s\,190}^3 n_{c\,4}^{-1/2} M_\odot)$ is reduced a
 factor of $10^{-4}$ from the initial value.
 
\section{DISCUSSION}

The size of the outflow reaches $z \sim 0.25 c_s/(4\pi G \rho_c)^{1/2}
 \sim 1800 c_{s\,190}n_{c\,4}^{-1/2}{\rm AU}$ in $\tau\sim 7000
 n_{c\,4}^{-1/2}{\rm yr}$.
However, CO observations show that the molecular outflows have typical
 spatial size of 8000 AU -- 4 pc (Fukui et al. 1993).
This simulation is limited in time and should be extened further.
However, calculation becomes harder after the core mass exceeds
 $\sim 0.1 c_{s\,190}^3 n_{c\,4}^{-1/2}M_\odot$ due to high density
 at the cloud center.  
Let us consider whether the magnetic angular momentum transport is
 effective still in the later stage $\tau > 7000 n_{c\,4}^{-1/2}{\rm yr}$.

As shown in Figure 1, $j$ is approximately proportional to $M$ at the
 epoch of the core formation.  
Writing down the relation as $j=p\times GM/c_s$, the coefficient $p$
 is approximately equal to 0.1 for the model shown in Figure 1.  
If $j$ is conserved, the centrifugal radius ($R_c$), at which the
 gravity is balanced with centrifugal force, is given by
 $R_c=j^2/GM=p^2GM/c_s^2\sim 250(p/0.1)^2(M/1M_\odot)c_{s\,190}^{- 2}{\rm AU}$
 [a similar relation is pointed out by Basu (1998)].  
This indicates that the centrifugal radius expands with time.  
On the other hand, a self-similar solution for the rotating isothermal
 thin disk has been found by Saigo \& Hanawa (1998).  
Their solution in the accretion phase also shows that the radius where
 $v_\phi$ takes the maximum increases with time as $\propto c_s \tau$.  
Since the outflow is ejected from a radius where $v_\phi$ is important
 (Paper I), these two indicate that the seeding region of the outflow
 moves outwardly.  
Using the solution by Saigo \& Hanawa (1998; for an example, see their
 Figure 7), the surface density ($\Sigma$) and thus the density ($\rho$)
 decrease as collapse proceeds and the seeding region moves outwards.  
This shows that the density near the seeding region decreases as the
 collapse proceeds.

This leads to the conclusion that the coupling between gas and
 magnetic fields (Nakano 1990) becomes stronger as long as we consider
 the seeding region, indicating that the mechanism of angular momentum
 transfer works also in the later stage of the evolution.

\acknowledgments 
The author thanks R. Ouyed, the referee, for his fruitful comments to improve
 the paper.
He also thanks T. Hanawa, T. Matsumoto, and F. Nakamura for useful discussion.
This work was partially supported by Grants-in-Aid from the Ministry
 of Education, Science, Culture, and Sports (10147105, 11640231).  
Numerical calculations were performed by Fujitsu VPP300/16 at the
 Astronomical Data Analysis Center, the National Astronomical Observatory,
 and NEC SX 4/2B at the Integrated Information Processing Center,
 Niigata University.

\clearpage \begin{figure}
\epsfxsize=\columnwidth\epsfbox{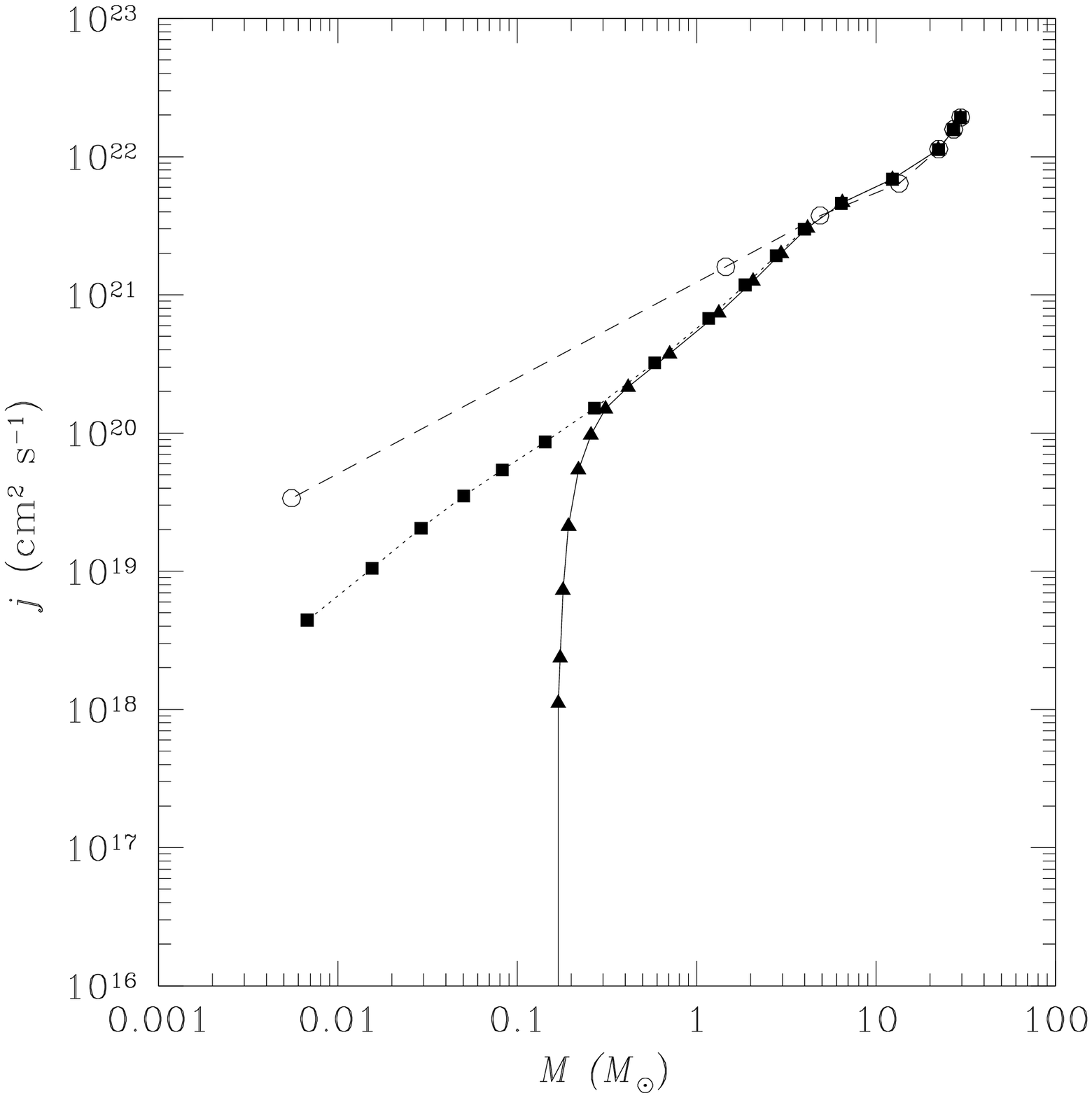} 
\caption{
Specific angular momentum $j$ is plotted against the accumulated
 mass from the center $M$.  
Open circles, filled squares, and triangles denote, respectively,
 the beginning and the end of the run-away collapse phase and the 
 accretion phase. 
\label{fig1} 
} 
\end{figure}

\begin{figure} 
\epsfxsize=\columnwidth\epsfbox{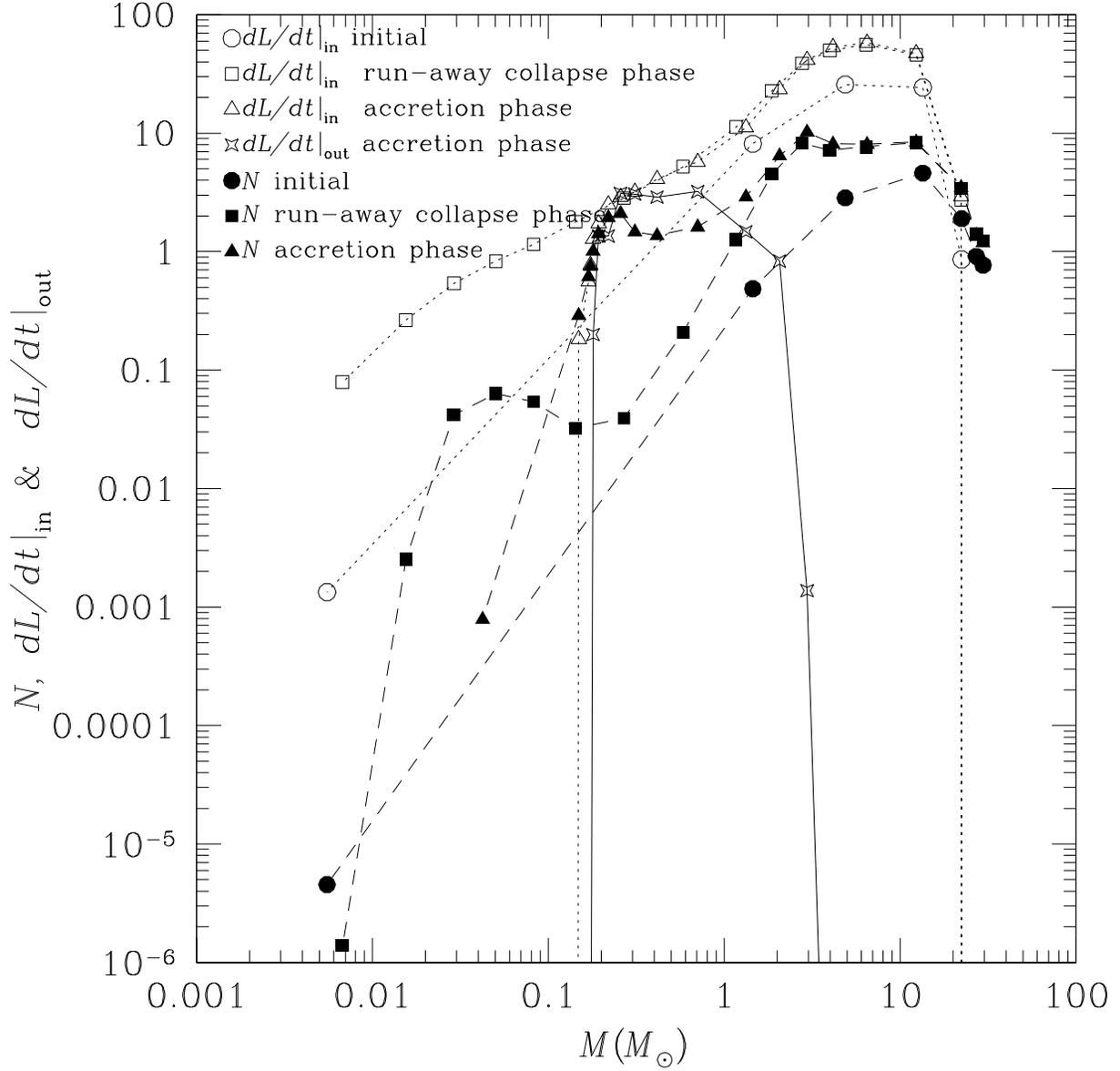} 
\caption{
The torque exerted on the cloud core by the magnetic fields, $N$,
 the angular momentum inflow rate convected by the inflow in the disk,
 $dL/dt|_{\rm in}$, and its outflow rate transferred by the outflow,
 $dL/dr|_{\rm out}$.  
Filled and open symbols denote respectively the distribution of $N$ and 
 that of  $dL/dt|_{\rm in}$.
The circles, and squares denote, respectively, the beginning
 and the end of the run-away collapse phase.
Open stars represent the angular momentum outflow rate $dL/dt|_{\rm out}$ in
 the accretion phase.
\label{fig2}
} 
\end{figure}
\end{document}